\documentclass[aps,preprint,floatfix]{revtex4}
\usepackage[dvips]{graphics,graphicx}
\usepackage{amsmath}
\begin{document}

\title{Quantum state of the fermionic carriers in a transport channel connecting particle reservoirs}
\author{Andrey~R.~Kolovsky$^{1,2}$ and Dmitrii~N.~Maksimov$^{1,3}$}
\affiliation{$^1$Kirensky Institute of Physics, 660036 Krasnoyarsk, Russia}
\affiliation{$^2$Siberian Federal University, 660041 Krasnoyarsk, Russia}
\affiliation{$^3$Siberian State Aerospace University, 660014 Krasnoyarsk, Russia}

\date{\today}
\begin{abstract}
We analyze quantum state of fermionic carriers in a transport channel attached to a particle reservoir. The analysis is done from the first principles by considering microscopic models of the reservoir and transport channel. In the case of infinite effective temperature of the reservoir we demonstrate a full agreement  between the results of straightforward numerical simulations of the system dynamics and solution of the specified master equation on the single-particle density matrix of the carriers in the channel. This allows us to predict the quantum state of carriers in the case where transport channel connects two reservoirs with different chemical potentials.  
\end{abstract}
\maketitle

\section{Introduction}

Electron transport in the mesoscopic devices is a wide subfield of the solid-state physics \cite{Datt95,Ferr97,Ihn09}. These studies are aimed to control the electron current between two or more contacts (electron reservoirs) attached to the device. Recently, the same kind of problems have been addressed for the principally different system -- neutral atoms in laser-based devices \cite{Bran12,Husm15,Krin15,Krin17} (see also theoretical works \cite{Brud12,Gutm12,Niet14,Pros14,Simp14,Chie14,Duja15,Kord15,112}). The main advantages of the latter system over the electron system is the perfect control over the system parameters and unique detection techniques that allow for {\em in situ} measurement of the quantum state of carriers in the device which, following Ref.~\cite{Bran12}, we refer to as the transport channel connecting particle reservoirs.

On the formal level the quantum state of carriers in the transport channel is characterized by the single-particle density matrix (SPDM), the knowledge of which suffices to predict the current between particle reservoirs. In the present paper we analyse SPDM of fermionic carriers from the first principles with the emphasis on decoherence effect of reservoirs. Clearly, to address this problem from the first principles one needs physically relevant microscopic models of the transport channel and particle reservoir. Having in mind cold atoms we model the transport channel by the tight-binding chain, which is known to adequately describe neutral atoms in deep optical lattices. (Here `deep' means that the width of the ground Bloch band is smaller than the energy gap separating it from the rest of the spectrum.) As concerns the particle reservoir, we model it by the Two-Body Random Interaction Model (TBRIM)  \cite{Bohi71,Fren71} that corresponds to a system of $N$ weakly interacting spinless fermions distributed over $M$ natural orbitals. The closed (isolated) TBRIM possesses the self-thermalization property \cite{izrailev,105} and it is shown in the recent work \cite{preprint} that self-thermalization is preserved if we open the system. This makes TBRIM an excellent model for the reservoir of fermionic particles.

The structure of the paper is as follows. After reviewing TBRIM in Sec.~\ref{sec2}, we attach the finite-length tight-binding chain to this reservoir and study particles propagation across the chain in Sec.~\ref{sec3}. We quantify decoherence effect of the reservoir on the carriers in the channel by the von Neumann entropy of SPDM and show that it is strictly positive. In Sec.~\ref{sec4} we compare the exact numerical results with those obtained by using the master equation on the reduced density matrix of the carriers (RDM).  Finally, in the concluding Sec.~\ref{sec5} we summarise the results and give the list of open problems.

\section{The model}
\label{sec2}
In this section we specify the system Hamiltonian $\widehat{H}$, which consists of the Hamiltonian of the particle reservoir $\widehat{H}_b$, the Hamiltonian of the transport channel $\widehat{H}_s$, and the coupling Hamiltonian $\widehat{H}_{int}$:
\begin{equation}
\label{a0}
\widehat{H}=\widehat{H}_b + \widehat{H}_s + \widehat{H}_{int} \;.
\end{equation}
%
\subsection{The particle reservoir}

We model the particle reservoir by TBRIM which describes $N$ interacting spinless fermions distributed over  $M$ natural orbitals with the energies $\epsilon_k$ ($\epsilon_{k+1} \geq \epsilon_k$):
\begin{eqnarray}
\label{a1}
\widehat{H}_b=\sum_{k=1}^M \epsilon_k \hat{d}_k^\dagger \hat{d}_k  
+\varepsilon_b \sum_{ijkl} V_{ij,kl}  \hat{d}_i^\dagger  \hat{d}_j^\dagger   \hat{d}_k \hat{d}_l \;. 
\end{eqnarray} 
Here operators $\hat{d}_i^\dagger, \hat{d}_i$ satisfy the usual anti-commutation relation and one-particle energies $\epsilon_k$ and  interaction constants $V_{ij,kl}$ are random (up to the obvious  symmetry relations insuring hermiticity of the Hamiltonian) variables with standard deviation equal to unity. The parameter $\varepsilon_b$ in the Hamiltonian (\ref{a1}) controls the strength of two-body interactions which couples every Fock state with other $K = 1+N(M-N)+N(N-1)(M-N)(M-N-1)/4$ Fock states. In the paper we assume $\varepsilon_b\ll 1$, i.e., we consider the limit of weakly interacting fermions. Yet, $\varepsilon_b$ is larger than some critical value where TBRIM shows the transition to Quantum Chaos \cite{Stock00,Haak10}. An analytical estimate for the critical  interaction strength can be obtained by using the {\AA}berg criteria \cite{Aber90}, while numerically this transition  is detected as the change of the level-spacing distribution from the Poisson distribution to the Wigner-Dyson distribution. In what follows we  fix the reservoir size to $M=12$, $N=6$, and set $\varepsilon_b=0.034$ where the energy level statistics perfectly follows the Wigner-Dyson distribution. The chosen $\varepsilon_b$ is approximately twice larger than the critical value. At the same time, it is small enough to speak about weakly-interacting fermions. In particular, the mean density of states, which in the case of non-interacting fermions is well approximated by the Gaussian of the width $\sim\sqrt{N}$, remains practically unaffected.  

Provided the condition of Quantum Chaos is satisfied, the system (\ref{a1}) shows the phenomenon of self-thermalization \cite{105}. It means that for any given eigenstate $|\psi_E\rangle$ occupation numbers of the natural orbitals $n_k=\langle \psi_E | \hat{d}_k^\dagger \hat{d}_k | \psi_E \rangle$  obey  (of course, with some fluctuations) the Fermi-Dirac distribution,
\begin{equation}
\label{a3}
n_k = \frac{1}{e^{\beta(\epsilon_k - \mu)} + 1}  \;,
\end{equation}
where the inverse effective temperature $\beta$ and the chemical potential $\mu$ are uniquely determined by the eigenstate energy $E$ and the number of particles $N$ through the solution of the following system of two non-linear algebraic equations,
\begin{equation}
\label{a4}
\sum_{k=1}^M\frac{1}{e^{\beta(\epsilon_k - \mu)} + 1} = N \;,\quad
\sum_{k=1}^M\frac{\epsilon_k}{e^{\beta(\epsilon_k - \mu)} + 1}  = E \;.
\end{equation}
Then  the ground and the highest energy eigenstates of the system (\ref{a1})  corresponds to $\beta=\pm\infty$, while an eigenstate from the middle of the spectrum  corresponds to $\beta=0$.  We mention, in passing,  that Eqs.~(\ref{a3}-\ref{a4}) also hold for the open TBRIM \cite{preprint}, where the number of particles changes in time.

\subsection{The transport channel}

We model the transport channel by the tight-binding chain,
\begin{eqnarray}
\label{a5}
\widehat{H}_s=V_g \sum_{l=1}^L  \hat{c}_l^\dagger \hat{c}_l 
 -\frac{J}{2}\left( \sum_{l=1}^{L-1}  \hat{c}_{l+1}^\dagger \hat{c}_l + h.c.  \right) \;,
\end{eqnarray} 
where $J$ is the hopping matrix element and $V_g$ has the meaning of the gate voltage. We shall characterise fermions in the channel by SPDM, 
\begin{eqnarray}
\label{a6}
\rho_{l,m}(t)=\langle \Psi(t) | \hat{c}_l^\dagger \hat{c}_m | \Psi(t) \rangle \;,
\end{eqnarray} 
where $| \Psi(t) \rangle$ is the total wave function of the whole system defined in the Hilbert space of the dimension
\begin{equation}
\label{a7}
{\cal N}=\frac{(M+L)!}{N!(M+L-N)!} \;.
\end{equation}
Since the quadratic form $\hat{c}_l^\dagger \hat{c}_m$ in Eq.~(\ref{a6}) conserves the number of particles,  the density matrix (\ref{a6}) can be presented as a sum of partial density matrices, 
\begin{eqnarray}
\label{a8}
\rho(t)=\sum_{i=1}^L \rho^{(i)}(t) \;,
\end{eqnarray} 
where $\rho^{(i)}(t)$ refer to the fixed number fermions in the channel. We note that for an isolated channel with $i$ fermions in a pure state the matrix $\rho^{(i)}(t)$ has $i$ eigenvalues equal to unity and $L-i$ eigenstates equal to zero.

\subsection{The coupling Hamiltonian}

Particles from the reservoir enter the transport channel by means of the coupling Hamiltonian 
\begin{eqnarray}
\label{a9}
\widehat{H}_{int}= \varepsilon \left( \sum_{k=1}^M W_k \hat{c}_{1}^\dagger \hat{d}_k + h.c.  \right) \;,
\end{eqnarray} 
where $W_k$ are random entries of the same magnitude as the interaction constants $V_{ij,kl}$  and $\varepsilon$ is our control parameter. In what follows we consider the situation where initially all particles are in the reservoir, i.e.,
\begin{eqnarray}
\label{a10}
| \Psi(t=0) \rangle = |\psi_E\rangle \otimes | vac \rangle \;.
\end{eqnarray} 
%

\section{System dynamics} 
\label{sec3}
In this section we discuss the system dynamics governed by the Schr\"odinger equation with the Hamiltonian (\ref{a0}) for the initial condition specified in Eq.~(\ref{a10}). 

\subsection{Population dynamics}

\begin{figure}
\includegraphics*[width=9.5cm]{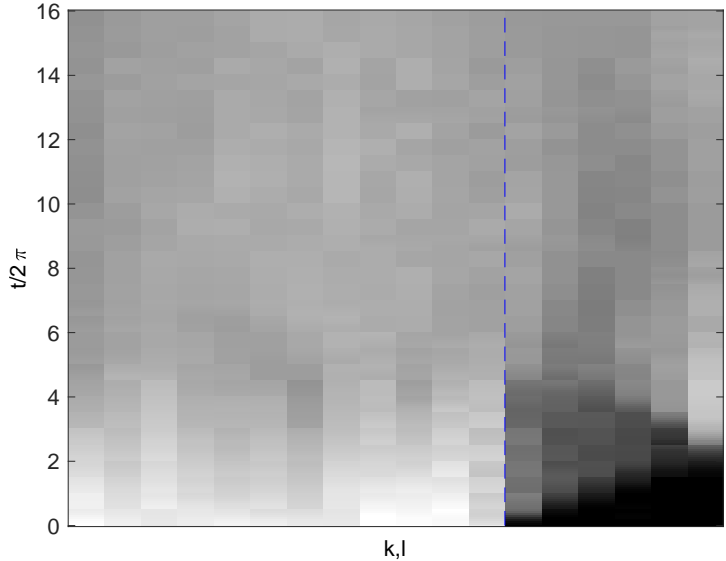}
\caption{Population dynamics of the reservoir orbitals (left to the vertical dashed line) and the lattice sites (right to the dashed line) for infinite  effective temperature of the reservoir.  The system size is $M=12$, $L=6$, and $N=6$. The value of the hopping matrix element $J=0.5$, the coupling constant $\varepsilon=0.1$. }
\label{fig1}
\end{figure}

Figure \ref{fig1} shows  occupation numbers  of the natural orbitals and the chain sites,
\begin{eqnarray}
\label{b1}
n_k(t)=|\langle \Psi(t) | \hat{d}_k^\dagger \hat{d}_k | \Psi(t) \rangle |^2  \;, \quad
n_l(t)=|\langle \Psi(t) | \hat{c}_l^\dagger \hat{c}_l | \Psi(t) \rangle |^2 \;,
\end{eqnarray} 
as the function of time for $|\psi_E\rangle$ from the middle of the spectrum of the system (\ref{a1}). (Thus we deal with the case of infinite temperature of the reservoir.) One distinguishes two qualitatively different stages/regimes in Fig.~\ref{fig1}. During the fist stage fermionic particles  propagate in the channel with the velocity determined by the hopping matrix element $J$ in Eq.~(\ref{a5}). Reaching the boundary particles are reflected back towards the reservoir. Notice that during this stage, which we refer to as {\em propagation stage}, the number of particles in the channel monotonically increases. During the second stage, which we refer to as {\em equilibration stage}, occupation of the chain sites and natural orbitals equilibrate at $n_k=n_l=N/(M+L)$.

\begin{figure}
\includegraphics*[width=9.5cm]{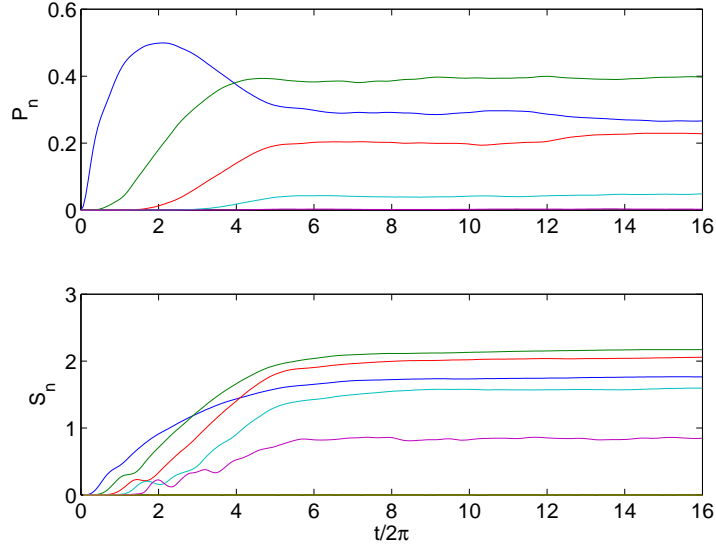}
\caption{Probabilities $P_i$ to find $i$ fermions in the transport channel  (upper panel) and von Neumann entropies $S_i$ of the normalized partial SPDMs (lower panel) as the functions of time.}
\label{fig2}
\end{figure}

To get a deeper insight into the population dynamics we calculate the partial density matrices $\rho ^{(i)}(t)$,  see Eq.~(\ref{a8}). The upper panel in Fig.~\ref{fig2} shows probabilities $P_i(t)$ to find $i$ fermions in the channel at a given time $t$, which is given by the equation
\begin{eqnarray}
\label{b2}
P_i(t)={\rm Tr}[\rho^{(i)}(t)]/i  \;.
\end{eqnarray} 
Increasing the evolution time we find $P_i(t)$ to approach the value $P_i(t=\infty)={\cal N}_i/{\cal N}$ where ${\cal N}_i$ is dimension of the sub-space of the Hilbert space defined by the condition that there are $i$ particles in the channel. This result proves that for infinite reservoir temperature we have complete equilibration between the system (the tight-binding chain) and the bath (TBRIM).

\subsection{Decoherence dynamics}

Next we discuss decoherence effect due to the reservoir. We characterize coherence of the carriers in the transport channel by the von Neumann entropy for the normalised partial SPDMs
\begin{eqnarray}
\label{b3}
\rho^{(i)}(t) \rightarrow  \frac{\rho^{(i)}(t)}{P_i(t)} \;, \quad
S_i(t)=-{\rm Tr}[\rho ^{(i)}(t)\log\rho^{(i)}(t)]  \;.
\end{eqnarray} 
Entropies $S_i(t)$ are depicted in the lower panel in Fig.~\ref{fig2}. It is seen that decoherence takes place immediately after the particles enter the transport channel and $S_i(t)$ quietly reach the maximally possible values  $\bar{S}_i=-i\log(i/L)$.  (The existence of this upper boundary is the main reason for considering the partial SPDMs, which refer to the fixed number of particles, instead of the total SPDM. In fact, the von Neumann entropy $S(t)=-{\rm Tr}[\rho(t)\log\rho(t)]$ of the total SPDM Eq.~(\ref{a5}) depends on the mean number of particles in the chain and, thus, an increase of $S(t)$ does not necessarily indicate decoherence.) Thus we conclude that in course of time every partial SPDM relaxes to a diagonal matrix proportional to the identity matrix, and so does the total SPDM.

\begin{figure}
\includegraphics*[width=9.5cm]{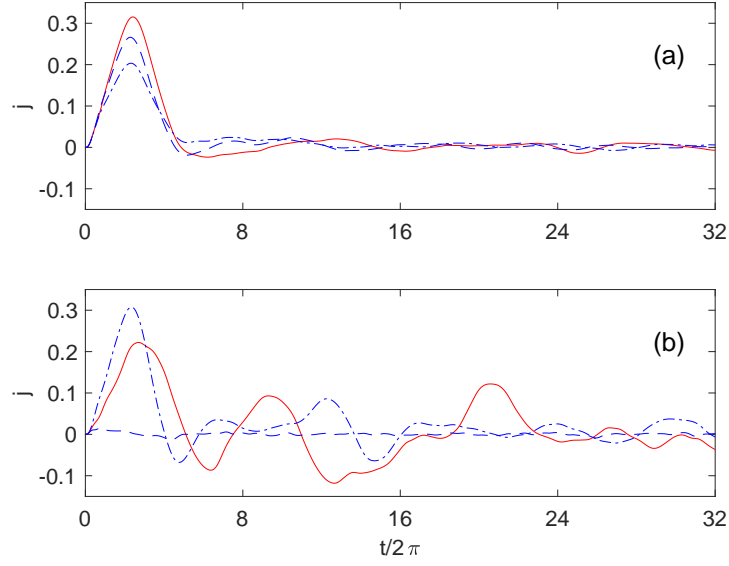}
\caption{ The total current in the transport channel as the function of time in the high-temperature (upper panel) and low-temperature (lower panel) limits. The dash-dotted, solid, and dashed lines correspond to different values of the gate voltage $V_g=-0.5, 0, 0.5$, respectively.}
\label{fig3}
\end{figure}

A direct consequence of the observed complete decoherence is an irreversible decay of the mean current $j(t)$,
\begin{eqnarray}
\label{b4}
j(t)={\rm Tr}[\hat{j}\rho(t)] \;,\quad  j_{l,m}=j_0\frac{\delta_{l,m-1}-\delta_{l-1,m}}{2i} \;,
\end{eqnarray} 
see red solid line in Fig.~\ref{fig3}(a).  We also mention that decay of the mean current is insensitive  (at least, on the qualitative level) to variation of the gate voltage $V_g$, see blue dashed and dash-dotted lines in Fig.~\ref{fig3}(a) which correspond to $V_g=\pm 0.5$. This is in a strong contrast with the low-temperature limit, where population dynamics and the mean current  crucially depend on inequality relation between the gate voltage and the Fermi energy $\epsilon_F$ (which is located at $\epsilon=0$ in the considered case of half-filling $N=M/2$). Indeed, in terms of the reservoir eigenstates the low-temperature limit corresponds to $|\psi_E \rangle$ close to the ground state, where occupation numbers $n_k$ of the natural orbitals show a pronounced step at $\epsilon_F$. Thus, fermions cannot enter the channel if $V_g>J$, where the whole conductance band lies above the Fermi energy. The results of numerical simulation of the low-temperature limit  fully confirm this expectation,  see Fig.~\ref{fig3}(b). Let us also notice the enhanced residual fluctuations of the current as compared to the  high-temperature limit. 

\section{Master equation approach}
\label{sec4}

It is interesting to compare the results of Sec.~\ref{sec3} with solution of the master equation on the reduced density matrix  ${\cal R}(t)={\rm Tr}_b[|\Psi(t)\rangle\langle\Psi(t)|]$ for fermionic carriers in the transport channel. Usually, one considers the following equation:
\begin{eqnarray}
\label{c1}
\frac{d {\cal R}}{dt}=-i[\widehat{H}_s,{\cal R}] -{\cal L}_{gain}({\cal R}) - {\cal L}_{loss}({\cal R}) \;, \\
\nonumber
{\cal L}_{loss}({\cal R})=\frac{\gamma}{2} (1-\bar{n})
(\hat{c}_1^\dagger\hat{c}_1{\cal R}-2\hat{c}_1{\cal R}\hat{c}_1^\dagger + {\cal R}\hat{c}_1^\dagger\hat{c}_1) \;, \\
\nonumber
{\cal L}_{gain}({\cal R})=\frac{\gamma}{2} \bar{n}
(\hat{c}_1\hat{c}_1^\dagger{\cal R}-2\hat{c}_1^\dagger{\cal R}\hat{c}_1 + {\cal R}\hat{c}_1\hat{c}_1^\dagger)  \;,
\end{eqnarray}
where $\bar{n}$ is the filling factor of the reservoir and $\gamma\sim \varepsilon^2$ is the relaxation constant.  (This equation also captures the case of bosonic carries, where the prefactor $(1-\bar{n})$ in the Lindblad  term ${\cal L}_{loss}$ should be replaced with  $(1+\bar{n})$ and fermionic annihilation and creation operators with bosonic operators.) It should be stressed that the standard derivation of the displayed master equation assumes a number of approximations \cite{Breu02,Dale14}, which have to be verified \cite{25,108}. In this sense Eq.~(\ref{c1}) implicitly refers to the high-temperature limit and is not valid in the low-temperature limit where, as it was demonstrated in the previous section,  the system dynamics depends on inequality relation between the Fermi energy and the gate voltage. (Notice that Eq.~(\ref{c1}) does not involve $\epsilon_F$ as a parameter.) For this reason from now on we focus on the high-temperature limit where all required assumptions/approximations are believed to be justified. 

\subsection{Populations and decoherence dynamics}

\begin{figure}
\includegraphics*[width=9.5cm]{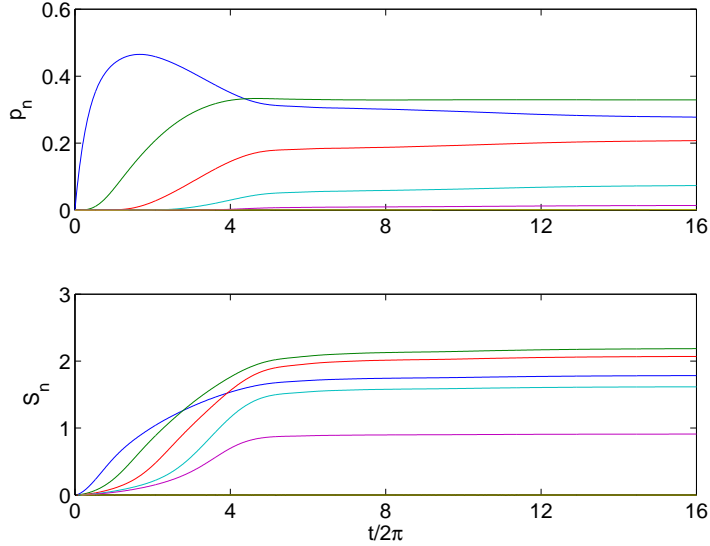}
\caption{The same quantities as in Fig.~\ref{fig2} yet calculated by using the master equation approach. The value of the relaxation constant is adjusted to $\gamma=0.4$.}
\label{fig4}
\end{figure}

It is easy to prove that the matrix ${\cal R}$  in Eq.~(\ref{c1}) has the block structure where each block is associated with the fixed number of fermions in the tight-binding chain of the length $L$. Using these blocks we calculate the partial SPDMs,
\begin{eqnarray}
\label{c2}
\rho^{(i)}_{l,m}(t)={\rm Tr}[{\cal R}^{(i)}(t) \hat{c}_l^\dagger\hat{c}_m ] \;,
\end{eqnarray}
and then use them to calculate probabilities $P_i(t)$ to find $i$ fermions in the transport channel and von Neumann entropies $S_i(t)$,  which characterize quantum state of these fermions. The results are presented in Fig.~\ref{fig4}, which should be compared with Fig.~\ref{fig2}. We notice that, when solving Eq.~(\ref{c1}), we take into account depletion of the reservoir, i.e., the parameter  $\bar{n}$ is decreased in time according to the depletion dynamics,
\begin{eqnarray}
\label{c3}
\bar{n}(t)=(N-N_s(t))/M \;.
\end{eqnarray} 
With this minor modification one finds very good agreement between the master equation approach and the exact numerical results. This agreement indicates that all assumptions/approximations used to derive Eq.~(\ref{c1}) are indeed justified.  This allows us to address within the framework of the master equation the more complex problem, where the transport channel connects two high-temperature reservoirs with different filling factors. 

\subsection{Stationary current between two reservoirs}

To take into account the second reservoir the master equation (\ref{c1}) on RDM should be complimented by two additional Lindblad terms which has the same structure as the Lindblad terms in Eq.~(\ref{c1}) but involves operators $\hat{c}_L^\dagger$ and  $\hat{c}_L$ instead of the operators $\hat{c}_1^\dagger$ and  $\hat{c}_1$. Also we redenote the parameters $\bar{n}$ and $\gamma$ as $\bar{n}_L$ and $\gamma_L$ (the left reservoir). Correspondently, the filling factor and relaxation constant of the right reservoir are denoted by $\bar{n}_R$ and $\gamma_R$ and,  to be certain, we assume $\bar{n}_L > \bar{n}_R$.

The solution of the described master equation with the source and sink terms was discuss in much details in Ref.~\cite{112} for the case of bosonic carriers.  Adopting the results of Ref.~\cite{112} to the currently considered case of fermionic carries we come to the following conclusions. In course of time SPDM relaxes to the three-diagonal matrix where (pure imaginary) off-diagonal elements of the matrix determine the stationary current $\bar{j}$ between the left and right reservoirs. This current is proportional to difference in the reservoir filling factors,  where the proportionality coefficient $A$ has particularly simple form in the case $\gamma_L=\gamma_R\equiv\gamma$,
\begin{eqnarray}
\label{c4}
A=\frac{1}{2}\frac{J\gamma}{J^2+\gamma^2} \;,
\end{eqnarray} 
and in the case $\gamma_R \ll \gamma_L\equiv\gamma$,
\begin{eqnarray}
\label{c5}
A=\frac{\gamma_R\gamma}{J^2+\gamma^2} \;.
\end{eqnarray} 
The latter case is of special interest for the purpose of microscopic analysis of the system dynamics. Indeed, if $\gamma_L\gg \gamma_R$ then the main source of decoherence is the left reservoir while the right reservoir barely serves as a particle sink. In the next subsection we analyze the problem where the left reservoir is modelled microscopically while the right reservoir is taken into account by using the master equation approach.

\subsection{Quasi-stationary current}

Following the discussion in the previous subsection, we consider the master equation
\begin{eqnarray}
\label{c6}
\frac{d {\cal R}}{dt}=-i[\widehat{H},{\cal R}] + {\cal L}_{loss}({\cal R}) \;, \\
\nonumber
{\cal L}_{loss}({\cal R})=\frac{\gamma_R}{2}
(\hat{c}_L^\dagger\hat{c}_L{\cal R}-2\hat{c}_L{\cal R}\hat{c}_L^\dagger + {\cal R}\hat{c}_L^\dagger\hat{c}_L) \;,
\end{eqnarray}
where $\widehat{H}$ is the Hamiltonian of the left reservoir with the attached transport channel. Due to large dimension  of the Hamiltonian we solve Eq.~(\ref{c6}) by using the stochastic approach \cite{Dale14}.  Specifically, we solve the Schr\"odinger equation of the form \cite{graham}
\begin{eqnarray}
\label{c7}
{\rm d} |\Psi\rangle= \left( -i\widehat{H}{\rm d}t - \frac{\gamma_R}{2} \hat{c}_L^\dagger\hat{c}_L  {\rm d}t
+\sqrt{\gamma_R} \hat{c}_L {\rm d}\xi \right) |\Psi\rangle \;,
\end{eqnarray} 
where ${\rm d}\xi$ is the Wiener process with $\overline{ {\rm d}\xi}=0$ and  $\overline{ {\rm d}\xi}^2={\rm d}t$.   Within this approach the reduced density matrix  ${\cal R}(t)$ is found by averaging the solution of Eq.~(\ref{c7}) over different realisations of the stochastic process, i.e.,  ${\cal R}(t)=\overline{ |\Psi(t)\rangle\langle \Psi(t) |}$.  The convergence of the averaging procedure is controlled against the condition  ${\rm Tr}[{\cal R}(t)]=1$.
\begin{figure}
\includegraphics*[width=9.5cm]{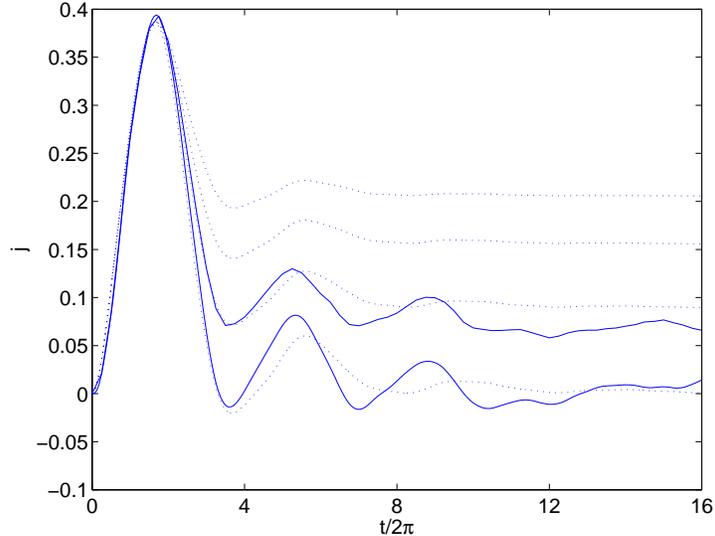}
\caption{The mean current in the transport channel connecting two reservoirs. Dotted lines show solution of the master equation with the source ($\bar{n}_L=0.5$, $\gamma_L=0.24$) and sink (${\bar n}_R=0$, $\gamma_R= 0,0.04,0.08,0.12$ from bottom to top) terms. The solid lines are solution of the master equation (\ref{c6}) for $\gamma_R=0$ and $\gamma_R=0.04$. }
\label{fig5}
\end{figure}

First we reproduce the result of Fig.~\ref{fig3}(a). The lower solid line in Fig.~\ref{fig5} shows the mean current in the transport channel for  $\varepsilon=0.1$ but slightly smaller system size $M=10$, $L=4$ (this reduces the dimension of the Hilbert space from  ${\cal N}=18564$ to ${\cal N}=3003$) and  $\gamma_R=0$. The exponential decay of the current is clearly seen. Next we set $\gamma_R$ to a small value $\gamma_R=0.04$. It is seen that $j(t)$ now decays to a finite value $\bar{j}$, i.e., we have a quasi-stationary current  between the reservoirs. We stress that  the observed rapid relaxation of the current to zero or finite value is exclusively due to decoherence effect of the left reservoir. In fact, matching the lower solid line to the solution of the master equation (\ref{c1}) we find $\gamma_L=0.24$. Thus we are indeed in the regime $\gamma_R \ll \gamma_L$ where one can neglect decoherence effect of the right reservoir.

\section{Conclusion}
\label{sec5}

We analyze quantum state of fermionic carriers in the transport channel connecting two reservoirs. The analysis is done from the first principles by considering a microscopic model of the reservoir (Two-Body Random Interaction Model) and the transport channel (tight-binding chain of a finite length). In the case of infinite effective temperature of the reservoirs the single-particle density matrix (SPDM) of  fermions in the channel is shown to relax to a three-diagonal matrix, whose off-diagonal elements determine the stationary current between the reservoirs. We stress that relaxation of SPDM  to this steady state is entirely due to decoherence effect of the reservoirs on the carriers propagating in the channel. We  obtain explicit expressions for the stationary current by justifying the master equation on the reduced density matrix of the carriers, which fortunately can be solved analytically.

The main challenge in the context of the presented studies is the case of low reservoir temperature, where occupation numbers  of its  natural orbitals show a step at the Fermi energy. It is believed that in this case the quantum state of fermionic carries in the transport channel is close to the Bloch wave with $k_F=\arccos[(\epsilon_F-V_g)/J]$.  In the other words, the stationary SPDM has many non-zero diagonals. In principle, one can prove or disprove this conjecture numerically within the framework of the discussed microscopic model by considering a larger system size [that will reduce residual fluctuations in Fig.~\ref{fig3}(b)]. The other route is a generalization of the master equation (\ref{c1}) onto the case of finite reservoir temperature, where it should include $k_F$ as an additional parameter. 


\newpage


\end{document}